\documentclass[twocolumn]{revtex4}

\usepackage{graphicx}
\usepackage{dcolumn}
\usepackage{bm}
\usepackage{gensymb}

\usepackage{amsmath, amssymb,amsfonts}
\usepackage{color}
\definecolor{red}{rgb}{1,0,0}
\definecolor{blue}{rgb}{0.3,0.3,0.9}

\begin{document}

\title{A Bright Spatially-Coherent Compact X-ray Synchrotron Source}

\author{S.~Kneip$^{1}$, C.~McGuffey$^{2}$, J.~L.~Martins$^{3}$, S.~F.~Martins$^{3}$, C.~Bellei$^{1}$, V.~Chvykov$^{2}$,
F.~Dollar$^{2}$, R.~Fonseca$^{3}$, C.~Huntington$^{2}$, G.~Kalintchenko$^{2}$, A.~Maksimchuk$^{2}$, S.P.D.~Mangles$^{1}$, T.~Matsuoka$^{2}$, S.~R.~Nagel$^{1}$, C.~Palmer$^{1}$, J.~Schreiber$^{1}$, K.~Ta Phuoc$^{4}$, A.G.R.~Thomas$^{2}$, V.~Yanovsky$^{2}$, L.~O.~Silva$^{3}$, K.~Krushelnick$^{2}$, Z.~Najmudin$^{1}$}

 \affiliation{$^{1}$ The Blackett Laboratory, Imperial College London, London, SW7 2BZ, UK}
 \affiliation{$^{2}$Center for Ultrafast Optical Science, University of Michigan, Ann Arbor, MI, 48109, USA}
 \affiliation{$^{3}$GoLP/Inst. Plasmas and Fus\~ao Nuclear, Instituto Superior T\'ecnico, Lisbon, Portugal}
 \affiliation{$^{4}$Laboratoire d'Optique Appliqu\'ee, ENSTA, Ecole Polytechnique, Palaiseau, 91761, France}

\begin{abstract}
Each successive generation of x-ray machines has opened up new frontiers in science, such as the first radiographs and the determination of the structure of DNA.
State-of-the-art x-ray sources can now produce coherent high brightness $>$~keV x-rays and promise a new revolution in imaging complex systems on nanometre and femtosecond scales.
Despite the demand, only a few dedicated synchrotron facilities exist worldwide, partially due the size and cost of conventional (accelerator) technology \cite{BilderbackJPhysB2005}.
Here we demonstrate the use of a recently developed compact laser-plasma accelerator \cite{ManglesSPD_Nature_2004,GeddesCGR_Nature_2004,FaureJ_Nature_2004} to produce a well-collimated, spatially-coherent, intrinsically ultrafast source of hard x-rays. This method reduces the size of the synchrotron source from the tens of metres to centimetre scale, accelerating and wiggling a high electron charge simultaneously. This leads to a narrow-energy spread electron beam \cite{ManglesSPD_Nature_2004,GeddesCGR_Nature_2004,FaureJ_Nature_2004} and x-ray source that is $>1000\times$ brighter than previously reported plasma wiggler \cite{RousseA_PRL_2004, KneipS_PRL_2008} and thus has the potential to facilitate a myriad of uses across the whole spectrum of light-source applications.
\end{abstract}

\maketitle

\section{Introduction}

There are many proposals to use extreme non-linear interactions of the latest generation of high-power ultrashort pulse laser systems to produce beams of high energy photons with high-brightness and short pulse duration.
For example, high-order harmonic generation promises trains of coherent pulselets \cite{DromeyB_NaturePhysics_2006,EsirkepovTZH_PRL_2009} whilst Compton scattering could extend energies into the $\gamma$-regime \cite{SchoenleinRW_Science_1996, KandoM_PRL_2007}. However, these and other similar schemes are complex and have stringent laser requirements.
An alternative proposal has been the use of compact laser-plasma accelerators to drive sources of undulating/wiggling radiation.

These accelerators use the plasma wakefield generated by the passage of an intense laser pulse through an underdense plasma \cite{TajimaT_PRL_1979}. Such wakefields can have intrinsic fields $> 1000\times$ greater than the best achievable by conventional accelerator technology, and thus can accelerate particles to high energies in a fraction of the distance. Recently, it has been demonstrated that at high laser power, the wakefield can be driven to sufficient amplitude that it can trap particles from the background plasma and accelerate them in a narrow energy spread beam\cite{ManglesSPD_Nature_2004, GeddesCGR_Nature_2004, FaureJ_Nature_2004}, now producing beams of electrons of  $\sim$ GeV energy on the order of 1 cm \cite{LeemansWP_NaturePhysics_2006, KneipS_PRL_2009}.

Such electron sources are clearly of interest to replace the accelerators which drive current synchrotron sources, and typically use multiple periods of alternately poled magnets (undulators or wigglers) to reinforce the synchrotron emission over a length of $\sim$ metres. The first demonstrations of wakefield driven radiation using external wigglers have now indeed been reported, though still being limited to optical or near-optical wavelengths and modest photon numbers  \cite{SchlenvoigtHP_NaturePhysics_2008, Fuchs2009}.

However, the particles being accelerated in the plasma accelerator also undergo transverse (betatron) oscillations due to the focusing fields of the plasma wave. The oscillations occur at the betatron frequency $\omega_{\beta}=\omega_{p}/\sqrt{2 \gamma}$, where $\omega_{p}$ is the plasma frequency and $\gamma$ is the Lorentz factor of the electron beam. Since this betatron wavelength ($\lambda_{\beta} \approx 2\pi c / \omega_{\beta}$) is much smaller than the period of comparable external wigglers, again due to the extremely large electric fields of the wakefield, the wavelength at which these particles radiate (of order $\lambda \sim \lambda_{\beta}/\gamma^{2} $) can be in the x-ray region for $\gamma \gtrsim 200$ i.e.~for an electron beam of energy $E > 100$ MeV. For small  transverse amplitude and thus small strength parameter  $K=\gamma r_{\beta} \omega_{\beta}/c \ll1$ (i.e.~undulator limit), the radiation spectrum will be narrowly peaked about a fundamental energy. 
As $K \rightarrow 1$, radiation also appears at harmonics. For large $K \gg 1$ (large amplitude - wiggler limit) a synchrotron spectrum with broad emission consisting of closely spaced harmonics is produced \cite{EsareyE_PRE_2002}. 
\begin{equation}
\frac{d^2I}{dE d\Omega} \Biggr|_{\theta=0} \simeq N_{\beta} \frac{6e^2}{\pi^2 c}\gamma_{z}^2\left(\frac{E}{E_{\text{crit}}}\right)^2\cdot \mathcal{K}^2_{2/3}\left(\frac{E}{E_{\text{crit}}}\right) 
\end{equation}
Here, $I$ is the radiated energy, $N_\beta$ is the number of oscillations and $E_{\text{crit}}=3\hbar K \gamma_z^2 \omega_{\beta}$ is twice the energy above and below which roughly half of the total energy is radiated. $\mathcal{K}_{2/3}$ is a modified Bessel function of the second kind. For $E > E_{crit}$, the spectrum of radiated energy decays exponentially. The radiation is confined to a cone with opening angle $\theta \approx K / \gamma_{z}$.

Previous measurements of betatron radiation from a broad energy spread electron beam from a laser wakefield accelerator have demonstrated x-ray emission with energies up to $\sim 1$ keV with a brightness up to $10^{19}$~ph/s/mrad$^2$/mm$^2$$/0.1\%$BW  \cite{RousseA_PRL_2004}. Here we demonstrate the use of a compact accelerator (in total $< 1$ cm), which produces a beam with much higher charge at high energy ($\sim 250$ MeV), to produce a 1-100 keV x-ray source with more than $1000 \times$ greater brightness than previous laser driven betatron sources \cite{RousseA_PRL_2004, KneipS_PRL_2008}. Furthermore, we demonstrate that this radiation source exhibits evidence for spatial coherence.

\section{Experimental Results}

The experiment was performed by focusing an intense short pulse ($\sim 30$ fs, $\sim 2$ J) laser onto the front edge of 3, 5 and 10 mm helium gas jets (see methods).
Electrons beams with narrow energy spread features were observed from all nozzles, at densities of $4-22\times10^{18}$ cm$^{-3}$.
Due to consecutive phases of injection \cite{KneipS_PRL_2009}, the electron beam consists of multiple filaments. For example, for a density of $8\times 10^{18}$ cm$^{-3}$ on the 5 mm nozzle, electron beams of $E = (230 \pm 70)$ MeV with $\Delta E = (25\pm10)$\% energy spread at FWHM were observed with an average of $2.2 \pm 0.4$ filaments per shot, with a root-mean-square (RMS) divergence of $1.5\times1.8$ mrad$^2$. 
The beam position had a RMS pointing fluctuation of $4.8 \times 4.7$ mrad$^2$.

The average and maximum energy of the electron beam follow the typical wakefield density scaling law \cite{LuW_PRSTAAB_2007,KneipS_PRL_2009}.

With the electron beam deflected away from laser axis by the spectrometer magnet, a bright (undeviated) beam of x-rays was also observed co-propagating along the laser axis. It was imperative to first prove that this x-ray source originates from the plasma itself.
In order to do this a grid of 60 $\mu$m diameter silver wires was placed a few centimetres from the target. X-rays originating from the interaction region project the outline of the mesh onto an imaging plate \cite{GalesSG_RSI_2004}. 
A strongly collimated beam of x-rays is evident in figure 1a. When either the laser power or plasma density was reduced to inhibit the electron beam, the x-ray beam also disappeared, showing that the generation of the x-rays is linked to the electron beam. The profile is elliptical with a FWHM divergence of $\theta_x \times \theta_y= 4\times13$ mrad$^2$, corresponding to a wiggler parameter $K=\theta \gamma$ of $K_x=1.5$ and $K_y=5$ for a simultaneously measured electron beam energy $E = 200$ MeV. The x-ray beam pointing is extremely stable, as can be deduced from figure 1b, which shows the sum of five consecutive shots. Their combined divergence is not significantly larger than that of a single shot measurement, the RMS pointing stability is 5 mrad in the horizontal and vertical direction, similar to the pointing stability of the electron beam.

To give an indication of the x-ray source size, microscopic objects were backlit with the x-ray beam. Figures 1c,g show photographs of collections of wires of various sizes, and figures 1 d,f,g,h show the corresponding radiographs of these objects. Even features as small as a 5 $\mu$m wire are resolved, indicating that the betatron x-ray source is $< 5 \, \mu$m, smaller than the size of the plasma wave in which the radiating electrons were trapped ($\approx 20\, \mu$m diameter).

\begin{figure}
\begin{center}
\includegraphics[width=8cm]{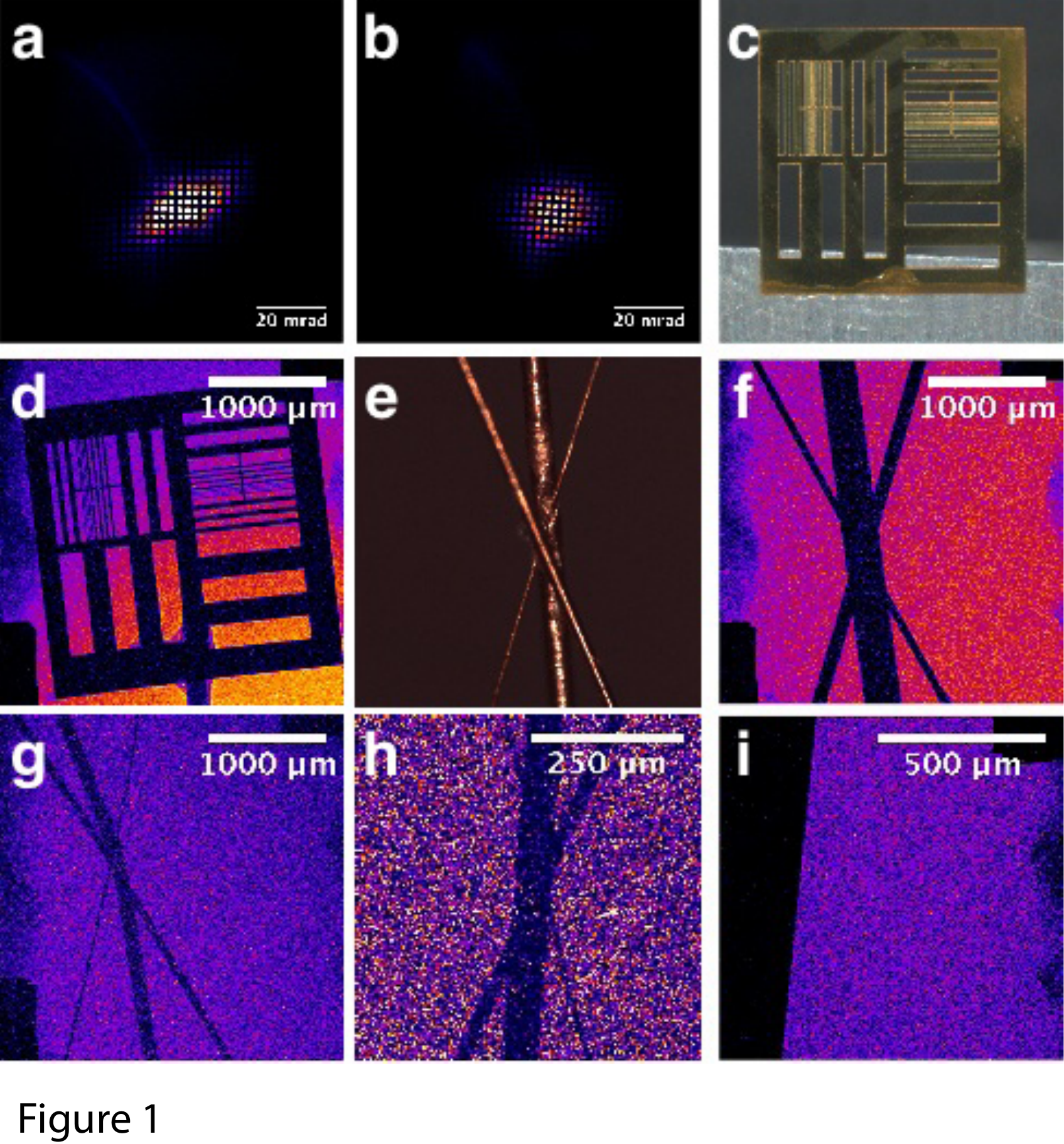}
\caption{X-ray beam profile behind a wire mesh from a single shot (a) and five consecutive shots (b) showing a $5\times11$ mrad$^2$ beam with $5$ mrad pointing stability. Photograph (c) and x-ray radiograph (d) of a 12 $\mu$m thick gold foil with rectangular 500 to 20 $\mu$m features. Photograph (e) and x-ray radiograph of wire triplets (f: 50/250/100 $\mu$m), (e,g: 50/100/20 $\mu$m), (h: 5/20/10 $\mu$m). All features are resolved indicating a betatron source $<5$ $\mu$m. (i) Single edge Fresnel diffraction from a cleaved InSb crystal produces a half-shadow, consistent with a $(2.0 \pm 0.5)$ $\mu$m spatially coherent source.}
\label{xraytargets}
\end{center}
\end{figure}

To quantify the source size more precisely, a half-plain was backlit with the x-ray beam. A typical intensity distribution on the detector looks like a half-shadow (figure 1i), whose details convolve information about the x-ray source and half-plain. The half-plain was a 0.25 mm thick cleaved InSb crystal ($<6\%$ transmission below 20 keV) and resembles an ideal step function. Assuming a  gaussian source function, convolution with the step function gives an error function which fits the experimental intensity distribution well. Figure 2 (inset) shows several experimental intensity lineouts and fitted error functions which predict the sharp rise and provide an estimate of the source size, in the range 2.9 - 6.0 $\mu$m at FWHM for the traces shown in figure 2 (inset).
The fit with the error function, however, fails to reproduce the ringing that follows the sharp rise on the high intensity side. 

To accurately model the shape of the intensity distribution, it is necessary to use Fresnel diffraction (see methods), where the details of the diffraction pattern depend on the spatial and spectral intensity distribution of the source and the dimensions of the setup. Figure 2 shows an experimental and modelled intensity distributions based on a gaussian spatial intensity profile and synchrotron spectrum, assuming a constant phase across the source. For the solid red curve in figure 2, $E_{\text{crit}}=8$ keV and an $1/e^2$ intensity radius of $w_x=1$ $\mu$m was assumed, which best reproduced both the sharp rise and the amplitude and width of the first fringe. 
Changing $w_x$ or $E_{crit}$ under or overestimates the height and/or width of the overshoot and/or the rise, as shown by the other curves in figure 2.
This determines that the $1/e^2$ radius of the source is in the range $0.5$ and $2$ $\mu$m and $E_{\text{crit}}$ between $4$ and $16$ keV.

The ringing must be due to interference of the radiation that has originated from different regions of a spatially extended source. If the phase relationship across the source was random, the fringes would be blurred, as shown in gray open circles in figure 2. The spectral width of the source also reduces fringe visibility. The fact that the data exhibits a clear first fringe can only be attributed to spatial coherence of the source. 

The x-ray yield measured with an x-ray CCD was studied as a function of electron charge. The x-ray signal increases with interaction length from 5 mm to 10 mm (Figure 3a). This is not surprising since a longer interaction length can accommodate more betatron oscillations. However, the x-ray yield per charge increases more than linearly from the 5 mm to 10 mm nozzle, by an average factor of 3. The electron beam exhibits oscillatory transverse deflections in the spectrometer normal to the dispersion direction, which is evidence for the presence of betatron oscillations during acceleration \cite{GlinecY_EPL_2008}. Indeed a high x-ray yield is correlated with the prominence of these betatron remnants (Figure 3a inset). An interaction of the accelerating electron beam with the laser (betatron resonance), can promote an increase in oscillation amplitude \cite{KneipS_PRL_2008}. For this to occur, the electron beam must catch up with the laser, which would be favoured by a longer interaction.
The faster than linear scaling of the x-ray yield may be a coherent effect based on micro-bunching, which simulations have shown as a consequence of a betatron resonance \cite{PukhovA_PoP_1999}. 

\begin{figure}
\begin{center}
\includegraphics[width=6cm]{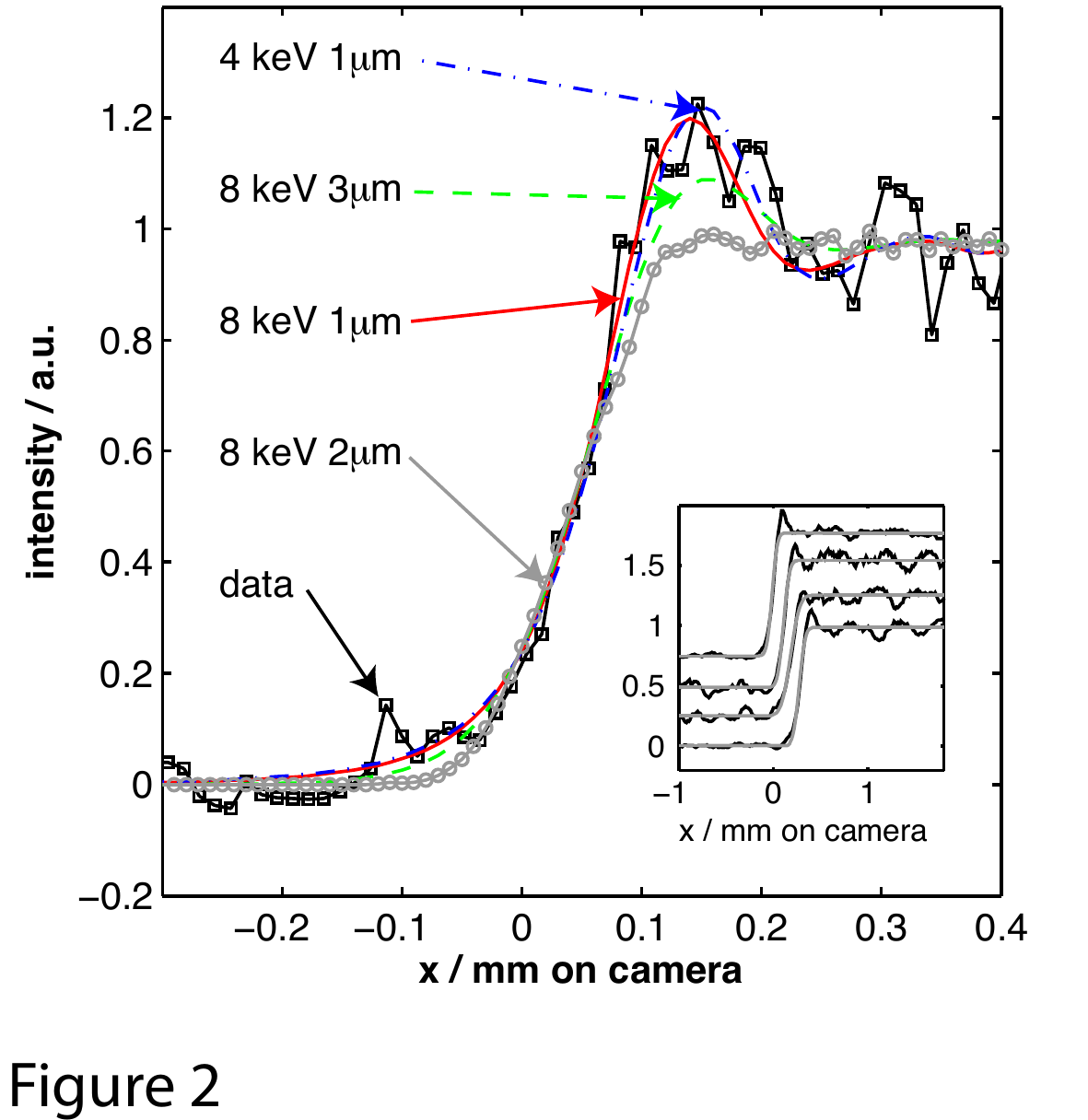}
\caption{The x-ray source casts a shadow of a half-plane on the detector. (inset) Intensity distribution of the half-shadow integrated along the edge (solid black) and fit with error functions (solid grey) for four different shots, giving source sizes of  2.9, 3.8, 6.0 and 3.9  $\mu$m at FWHM from top to bottom, assuming gaussian source distribution. Error functions mimic the observed intensity distribution closely but fail to reproduce the oscillations. (main)  Close-up of measured (open black squares) and modelled intensity distribution using Fresnel diffraction methods, assuming a spatially coherent gaussian source with synchrotron spectrum $E_{\text{crit}}$/$w_{0}$ of 8 keV/1 $\mu$m (solid red), 8 keV/3 $\mu$m (dashed green) and 4 keV/1 $\mu$m (dash-dotted blue) and an incoherent source with 8 keV/2 $\mu$m (open grey cicles).
}
\label{sourcesize}
\end{center}
\end{figure}

The spectral properties of the betatron radiation were determined by measuring the x-ray transmission through a set of filters with an x-ray CCD detector \cite{KneipS_PRL_2008}. 
Assuming the spectrum is synchrotron-like, the measured $E_{\text{crit}}$ increases with density and electron energy. 
For the 5 mm nozzle at $n_e=(1.0\pm0.4)\times 10^{19}$ cm$^{-3}$, the measured $E_{\text{crit}}=(29\pm13)$ keV as shown in figure 3b.

\begin{figure}
\begin{center}
\includegraphics[width=6cm]{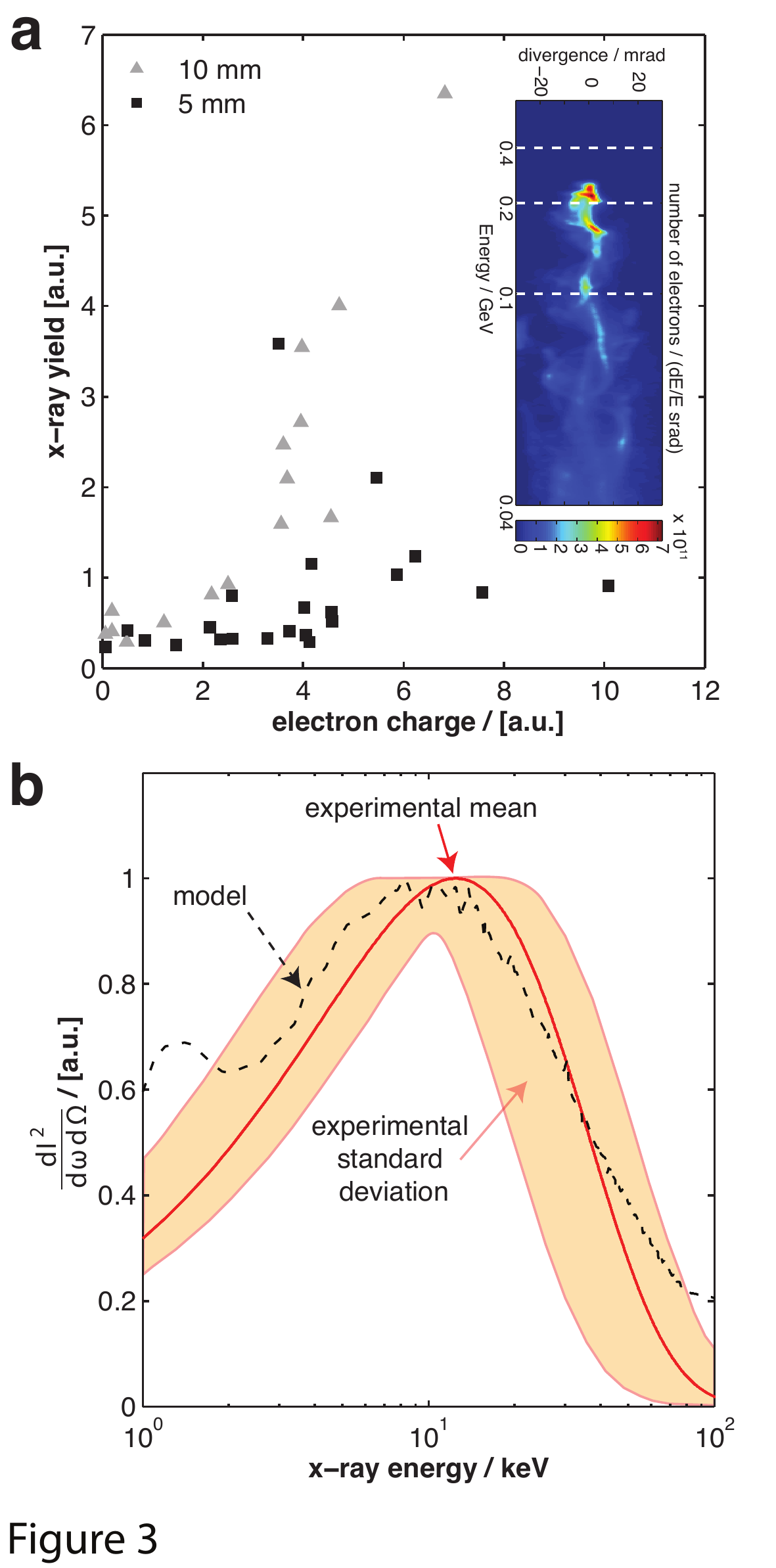}
\caption{(a) X-ray yield as a function of electron charge, for the 5 mm (black squares) and 10 mm (grey triangles) nozzle. The x-ray signal increases faster than linearly with increasing interaction length (and thus $N_{\beta}$).  (a inset) shows an energy-dispersed electron spectrum with prominent transverse features remnants of betatron oscillations. (b) The betatron spectrum obtained from numerical modelling agrees well with the experimentally measured spectrum within one standard deviation.}
\end{center}
\label{energyscaling}
\end{figure}

\section{Numerical Simulations}
Numerical modelling was carried out using electron trajectories obtained from the fully relativistic PIC code OSIRIS. The simulations were run in the boosted frame, which due to relativistic length contraction and time dilation, offer higher resolutions for shorter run times. The trajectories were post-processed to yield the characteristics of the betatron radiation (see methods). 
Figure 4a shows a simulation for a 5 mm nozzle at $n_{e} = 8\times10^{18}$~cm$^{-3}$, the spectrally integrated x-ray beam profile with a divergence of $5\times16$ mrad$^2$. The profile is elongated in the direction of laser polarisation, demonstrating excellent quantitative agreement with the measurement (figure 1a). Figure 4b shows the modelled x-ray spectrum as a function of energy and angle. The x-ray flux peaks on-axis at 10 keV and extends to $\simeq100$~keV. A lineout taken on-axis, as would be measured by our detector, is plotted in figure 3b, comparing well with the measured spectrum.
A total of $10^8$ photons are predicted between 1 and 84 keV from a 30 fs electron bunch, where we measure $10^6$ to $10^8$, depending on electron charge, corresponding to a peak brightness $10^{21}-10^{23}$ ph/s/mrad$^2$/mm$^2$/0.1\%BW. The x-ray pulse duration is of the same order as the electron bunch duration, and is thus of $\sim$fs timescale. The modelling reveals that the fastest electrons oscillate with the smallest amplitudes $r_{\beta} \lesssim 2 \mu$m, consistent with the measured source sizes. K-parameters range from 1-30 with an average of $\simeq7$ close to the measured experimental value.

\begin{figure}
\begin{center}
\includegraphics[width=8cm]{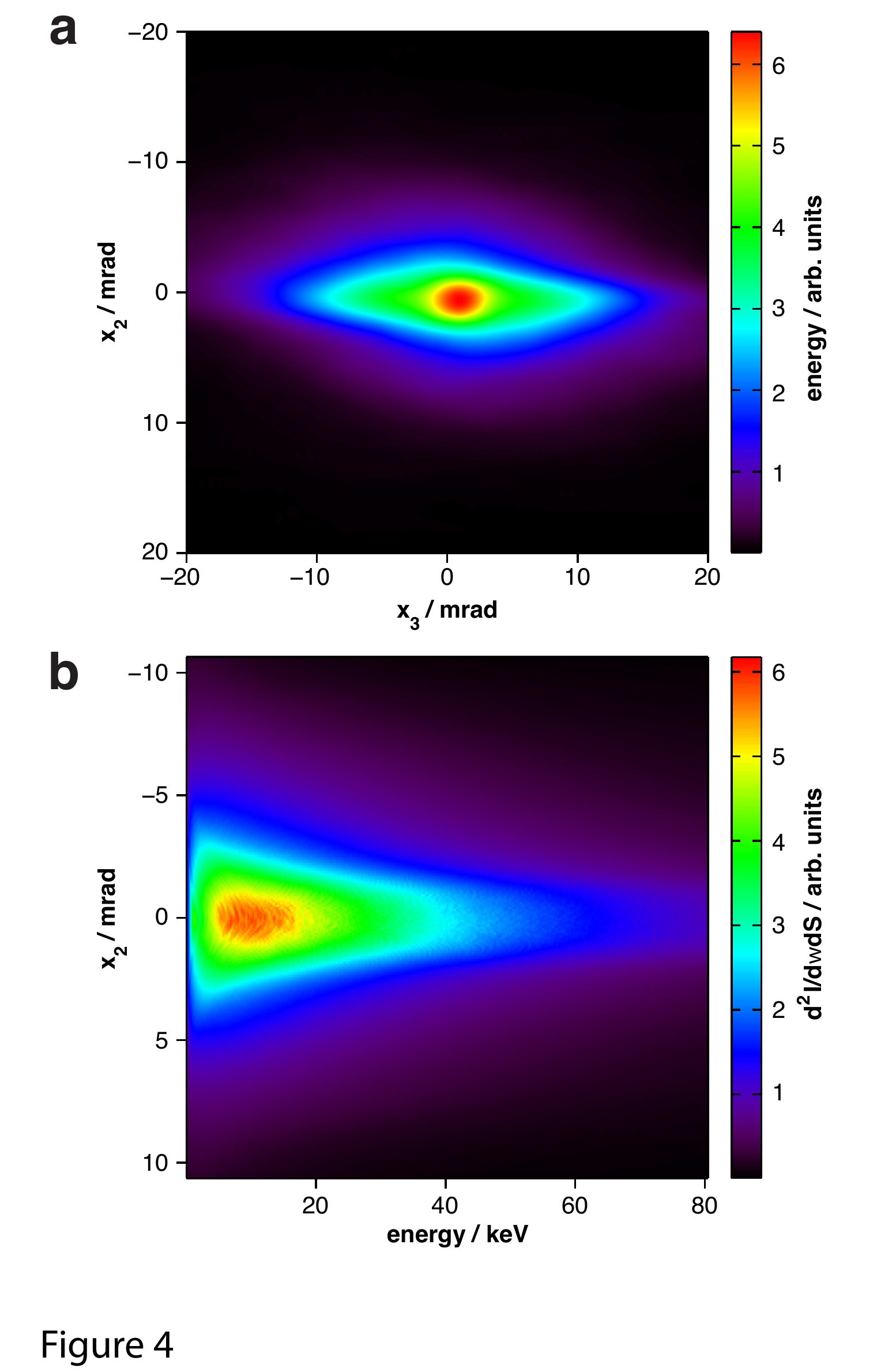}
\caption{Numerical modelling of interaction; (a) spectrally integrated x-ray beam profile showing an elliptically elongated beam profile in the direction of laser polarisation, (b) angularly and spectrally resolved x-ray flux displaying a peak on axis at $10$ keV and a tail extending to $\simeq100$ keV.}
\end{center}
\label{energyscaling}
\end{figure}

\section{Conclusions}

In conclusion, we have shown, that a laser driven plasma can simultaneously serve as particle accelerator and wiggler, producing high quality beams of x-rays, with mrad divergence, $\mu$m source size, tens of keV critical energy and peak brightness $\sim 10^{23}$~ph/s/mrad$^2$/mm$^2$/0.1\%BW. 
The measured radiation is bright, spatially-coherent, and intrinisically ultrafast opening up a multitude of advanced applications, such as phase contrast and lensless imaging, previously only possible with large conventional light sources \cite{NeutzeR_Nature_2000,WilkinsSW_Nature_1996}. The laser plasma wiggler has the potential of making high-brightness radiation sources widespread, and thus impacting all areas of science and technology.

 

\section{Methods}

\subsection{Laser}
The experiments were carried out on the Hercules laser at the University of Michigan. Linearly polarised pulses with a central wavelength of $\lambda_{0}=800$ nm, a gaussian full-width half maximum (FWHM) pulse duration of $\tau=32$~fs and a maximum energy of 2.3 J were focused to a peak intensity of $I=4.7\times 10^{19}\, {\rm Wcm}^{-2}$ or normalised vector potential $a_0=4.7$ with an off-axis parabolic mirror of focal length $f=1$ m and $F$-number of $F=10$. The transverse intensity profile of the laser in vacuum yields a focal spot diameter of $d_{\mathrm{FWHM}}=11.2$ $\mu$m.
Typically $55$~$\%$ of the pulse energy is within $d_{\mathrm{FWHM}}$.

\subsection{Fresnel Diffraction}
Fresnel diffraction occurs when a wave passes through an aperture and diffracts in the near field, causing any diffraction pattern observed to differ in size and shape, depending on the distance between the aperture and the projection $i$, the size of the aperture $A$ and the wavelength of the wave $\lambda$. In the present case, $i\simeq1500$ mm, $\lambda < 5\times10^{-7}$ mm ($>2$ keV) and $a\simeq125$ $\mu$m assuming an x-ray beam of 5 mrad divergence and a half-plain aperture at $o=50$ mm from the source. Therefore the Fresnel number $F=A^2/(i\lambda) \approx 16>1$, requiring a treatment in the Fresnel regime.
The intensity distribution of a monochromatic point source on the projection is given by
\begin{equation}
I(x)=E\cdot E^{\ast} = \frac{I_0}{2} \left\{ \left( \frac{1}{2} + \mathcal{C}(w(x)) \right)^2 + \left( \frac{1}{2} + \mathcal{S}(w(x)) \right)^2 \right\}
\end{equation}
where $E$ and $E^{\ast}$ are the electric field and its complex conjugate, $\mathcal{C}$ and $\mathcal{S}$ are the Fresnel functions, $w(x)=x \left(o/(i+o)\sqrt{2(1/i+1/o)/\lambda}\right)$ and $x$ is the position on the projection.

If the source is not point-like, the electric field has to be convolved with the source function and spectral distribution $E(\lambda,x)=\int_{y} \int_{\lambda} R(\lambda)g(y) E(\lambda,x-i/o \cdot y)\, \text{d}y \,\text{d}\lambda$ to compute $I(x)$, where $R(\lambda)$ is the spectral response of the detector.

\subsection{Numerical Modelling}
Three-dimensional numerical simulations were performed with the particle in cell code OSIRIS \cite{FonsecaRA_LNICS_2002}, in which a linearly polarised pulse with experimental parameters is focused 0.25~mm into the plasma. 
The plasma density profile increases linearly from zero to $n_e = 8\times 10^{18}~$cm$^{-3}$ in the first 0.5~mm, it is constant for 3.5~mm, and falls linearly to zero in 0.5~mm. 
Simulations were performed in a relativistically boosted frame ($\gamma = 5$), which allows for significant computational gains \cite{VayJL_PRL_2007}. The simulation box corresponds to $60\times94\times94~\mu$m$^3$ in the laboratory frame. 
A total of $9.2\times10^8$ particles were pushed for $5\times 10^3$ iterations. The resolution in the laser propagation direction $z$ is $k_0\Delta z  = 0.11$, and $k_p\Delta x = k_p\Delta y = 0.16$ in the transverse directions.
To obtain the radiation emission from the simulation, a post-processing diagnostic is used on a set of trajectories of injected electrons \cite{MartinsJL_SPIEE_2009}. The position and momentum history of the electrons is used to deposit the radiated fields on a virtual detector.







\end{document}